# Policy Expectation Counts? The Impact of China's Delayed Retirement Announcement on Urban Households Savings Rates

## By Shun Zhang

This article examines the impact of China's delayed retirement announcement on households' savings behavior using data from China Family Panel Studies (CFPS). The article finds that treated households, on average, experience an 8% increase in savings rates as a result of the policy announcement. This estimation is both significant and robust. Different types of households exhibit varying degrees of responsiveness to the policy announcement, with higher-income households showing a greater impact. The increase in household savings can be attributed to negative perceptions about future pension income.

On November 12, 2013, a delayed retirement announcement was made in China, creating a unique opportunity to study how policy expectations influence households' daily decision-making. The announcement introduces a significant change in the retirement policy status and affects individuals' planning and savings behavior. Understanding the impact of such policy announcements on households is crucial for policymakers. In this study, I use China Family Panel Studies (CFPS) to build a comprehensive household-level dataset to examine the effects of the delayed retirement announcement on households' savings behavior.

To identify the causal impact of the announcement, the study adopts a Difference-in-Differences (DID) strategy, which has gained popularity in the literature. The key assumption underlying the DID strategy is the parallel-trend assumption, which means that in the absence of the treatment (delayed retirement announcement), the treated and control groups would have followed similar trends over time. By comparing the savings behavior of treated households (those affected by the delayed retirement announcement) with that of a control group (all retired households), we can isolate the specific impact of the policy change.

The findings of this study contribute to three strands of literature. Firstly, it provides insights into the role of policy expectations in shaping households' decision-making processes. By examining how the delayed retirement announcement influences households' savings behavior, we gain a deeper understanding of how policy information and expectations can drive economic decisions. This contributes to the broader field of policy studies.

Secondly, the study sheds light on the effects of retirement announcements on savings behavior. Retirement planning is a crucial aspect of financial security, and policy changes can significantly impact individuals' savings decisions. By analyzing the savings rates of treated households before and after the delayed retirement announcement, we can observe any shifts in behavior and explore the underlying reasons behind these changes. These insights are relevant for policymakers aiming to design effective retirement policies that align with households' financial needs and preferences.

Lastly, we can establish a causal relationship between the delayed retirement announcement and households' savings behavior by applying DID. This rigorous identification strategy helps overcome potential confounding factors and strengthens the credibility of the study's findings.

The remainder of this article is organized as follows. Section I briefly reviews the policy background. In Section II, I introduce the dataset and the empirical strategy. Section III reports the main results about how the delayed retirement announcement affected the savings rate of treated households. Section IV discusses the heterogeneity and robustness test. Section V presents a mechanism analysis. Section VI concludes.

## I. Policy Background

The delayed retirement policy entered the public spotlight in 2013 when, on November 12, the Third Plenary Session of the 18th Central Committee of the Communist Party of China approved the "Decision of the Central Committee of the Communist Party of China on Some Major Issues Concerning Comprehensively Deepening the Reform." This decision included a commitment to study and formulate the policy of progressively delaying the retirement age.[1] Consequently, the concept of delayed retirement garnered significant attention from the public.

As evidenced by the "Baidu Search Index," the keyword "delayed retirement" ("延迟退休" in Chinese) received unprecedented attention in November 2013. This phenomenon indicates not only the widespread public interest in this topic but also proves it as an exogenous announcement. Therefore, it is feasible for this study to employ the DID method to estimate the effects of the policy announcement on household savings rates.

---

[1] For reference: https://www.gov.cn/jrzg/2013-11/15/content_2528179.htm, 《中共中央关于全面深化改革若干重大问题的决定》

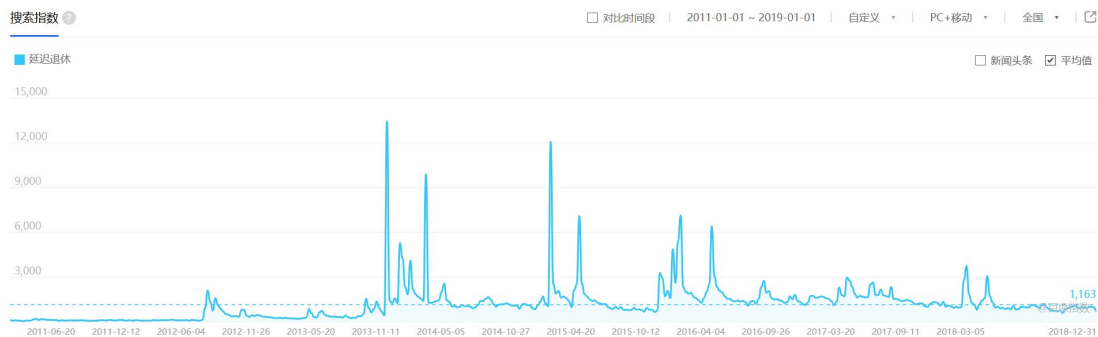

Fig. 1 Baidu Search index for "delayed retirement" keyword (2011-2018)
**Notes:** The figure was obtained on June 11, 2023, from https://index.baidu.com/v2/index.html#/, where you can find other indexes for the keywords the article mentioned.

Since then, the delayed retirement policy has been mentioned multiple times in CPC and Chinese government documents. For instance, the 20th CPC Congress report emphasized the need to improve the nationwide coordinated system for basic pension insurance, develop a multi-tiered and multi-pillar pension insurance system, and implement a progressive increase in the legal retirement age. However, as of June 2023, no specific policies regarding "delayed retirement" have been announced. Various detailed policy guessing, such as the rumored "progressive delayed retirement plan that both men and women retire at the age of 65 around 2055," has been denied by several local social security authorities.[2]

Therefore, the anticipated impact of the delayed retirement policy, as studied in this article, holds practical value and can serve as a benchmark for future policy implementation and an evaluation of policy effectiveness.

## II. Data and Empirical Strategy

### A. Data

The data utilized in this study is from CFPS, covering the years 2010, 2012, 2014, 2016, and 2018. Within each year's dataset, some variables of the household economic situation are kept, including information on household income, various expenditures, and wealth levels. Additionally, the individual database is merged by using the variable "pid." To ensure data consistency and facilitate the application of the DID, only household samples that were consistently surveyed from 2010 to 2018 are retained. Furthermore, the analysis is restricted to households residing in urban areas, since urban residents are more likely to be influenced by retirement policy.

The dependent variable analyzed in this article is the household savings rates. Referring to the method employed by Ma & Zhou (2014) and Can et al. (2019), two kinds of household savings rates are used in this article. Savings rate 1 is computed as the ratio of household savings to total household income: ((family income – family expenditure) / family income). Savings rate 2 is inspired by the hypothesis (e.g., Liu

---

[2] For reference: https://www.bbc.com/zhongwen/simp/chinese-news-64648455

2022) that a delayed retirement policy will influence individuals' expectations of future medical expenses and it is calculated as: ((family income – family expenditure + family healthcare expenditure) / family income). I removed the samples which on the top or the bottom 5% of the savings rates in the database. Furthermore, it is worth emphasizing that many households have negative savings rates, which may be attributed to the calculation method, as it has an upper bound of 1 but no lower bound.

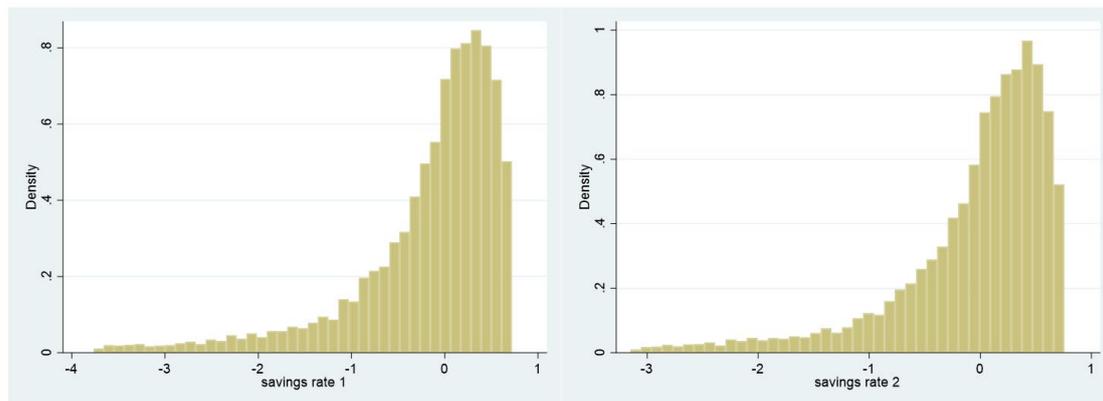

Fig. 2 Histogram of household savings rate

**Notes:** This figure shows how the dependent variable was distributed. The left one uses the savings rate that is directly calculated from the ratio of saving to income. The right one uses the adjusted saving (adding healthcare expenditure).

The identification of the treatment group and control group in this study is based on the retirement status of household members in the first year of the policy announcement (2014). If all members retired in 2014, the household is classified as the control group (17% of total samples), unaffected by the expected impact of the delayed retirement policy. Conversely, if at least one member does not retire in 2014, the household is categorized as the treatment group (83% of total samples). The retirement policy in China is mandatory, and most residents do not have the option to choose their retirement status individually. Therefore, the group assignment can be considered exogenous. Following the approach of Moser & Voena (2012) and Can et al. (2019), the study further considers the intensity of treatment by dividing the treatment group households based on the number of labor force members in 2014. This measure serves as an indicator of the expected policy treatment effect. Therefore, the subsequent analysis can employ a continuous form of DID model.

Moreover, the study controls household-level covariates, including logarithm of total household income, logarithm of household savings, logarithm of total assets, house value (measured in RMB Yuan, in tens of thousands), household population size, the proportion of household members hospitalized, and the proportion of household members retired.

Table 1 provides summary statistics of covariates and dependent variables for the treatment and control groups in our sample. From Panel A, we can find the control group is not directly comparable with the treatment group along several key dimensions. For example, the control group on average has a considerably higher savings and lower family size than the treatment group. This suggests that the treatment group is less wealthy and has lower family member average age. Thankfully, DID does not require covariates to be balanced between the treatment and the control. From Panel B, we can further find the control households on average save more, consistent with the common sense that the old are more thrift.

Table 1 Summary Statistics

|  | Treatment group | | Control group | | |
| --- | --- | --- | --- | --- | --- |
|  | Mean | SD | Mean | SD | Diff. |
|  | (1) | (2) | (3) | (4) | (5) |
| Panel A. Covariates of the treatment and control group | | | | | |
| logarithm of household income | 10.71 | 0.0151 | 10.49 | 0.0395 | 0.221*** |
| logarithm of household savings | 6.907 | 0.0561 | 7.457 | 0.133 | -0.551*** |
| logarithm of total assets | 12.62 | 0.0227 | 12.5 | 0.0585 | 0.120*** |
| house value | 48.84 | 1.276 | 56.69 | 3.109 | -7.842*** |
| household population size | 3.88 | 0.0274 | 2.343 | 0.0403 | 1.536*** |
| hospitalization rate | 0.263 | 0.00566 | 0.358 | 0.0141 | -0.0948*** |
| retirement rate | 0.137 | 0.00315 | 0.625 | 0.00694 | -0.488*** |
| Panel B. Dependent variables of the treatment and control group | | | | | |
| savings rate 1 | -0.191 | 0.00851 | -0.137 | 0.0181 | -0.0536** |
| savings rate 2 | -0.0722 | -0.00746 | 0.038 | 0.0160 | -0.111*** |

**Notes:** The table reports the statistics of the treatment and control samples. The control sample consists of households that all the adult members have retired (identified by whether receiving a pension) and the treatment sample consists of households with a least one working member. Standard errors are clustered at the household level. *** P<0.01, ** P<0.05, and * P<0.1.

### B. Empirical Strategy

In this study, I construct the following model to examine the direct impact of the delayed retirement announcement on the household savings rate:

$$Y_{it} = \gamma_i + \lambda_t + \delta(Treat_i \cdot Post_t) + X'_{it}\beta + \varepsilon_{it} \quad (1)$$

$Y_{it}$ is the dependent variable and will be replaced by two measurements of household savings rate. $\gamma_i$ and $\lambda_t$ are household and year fixed effects $Treat_i \cdot Post_t$ is the DID variable and $\delta$ is the primary parameter of interest. $X'_{it}$ comprises household economic conditions and family members' employment and health status.

Considering that the effect of delayed retirement announcement on workers cannot be achieved overnight, and people's expectation of delayed retirement policy will also change over time. Therefore, there may be a dynamic effect of policy announcement on household saving rates. For this reason, I rewrite equation (1) as following:

$$Y_{it} = \gamma_i + \lambda_t + \sum_{\tau=0}^{m} \delta_{-\tau} D_{s,t-\tau} + \sum_{\tau=1}^{q} \delta_{+\tau} D_{s,t+\tau} + X'_{it}\beta + \varepsilon_{it} \quad (2)$$

$D_{s,t-\tau}$ and $D_{s,t+\tau}$ are the interaction term for the treatment dummy and year dummy, and their coefficients represent the dynamic effect of interest in this study. $X'_{it}$ represents covariates, consistent with equation (1).

Inspired by Chen et al. (2020), this study also considers the continuous DID approach, as demonstrated in Equation (3). $\%FW_i$ is the density of exposing policy announcements in household i, which is calculated as the ratio of working adults to total adults in 2014.

$$Y_{it} = \gamma_i + \lambda_t + \delta(\%FW_i \cdot Post_t) + X'_{it}\beta + \varepsilon_{it} \quad (3)$$

For the same reason, $D_{s,t-\tau}$ and $D_{s,t+\tau}$ in equation (2) can also be interpreted as the product of $\%FW_i$ and year dummy. As a consequence, I could identify the dynamic effects of different treatment intensities.

## III. Main Results

### A. Parallel Trend Test

The basic assumption of the DID is the parallel trend. It states that in the absence of policy shock, the trend of the outcome (dependent) variable should not exhibit systematic differences between the treatment group and the control group. While we cannot directly observe the counterfactual situation, we can test whether the parallel trend exists before the policy shock. Figure 3 presents the trend in household savings rates for the treatment group and the control group. It can be observed that before the announcement of the delayed retirement policy in 2013, the trends in both the treatment and the control were relatively similar, satisfying the parallel trend assumption.[3]

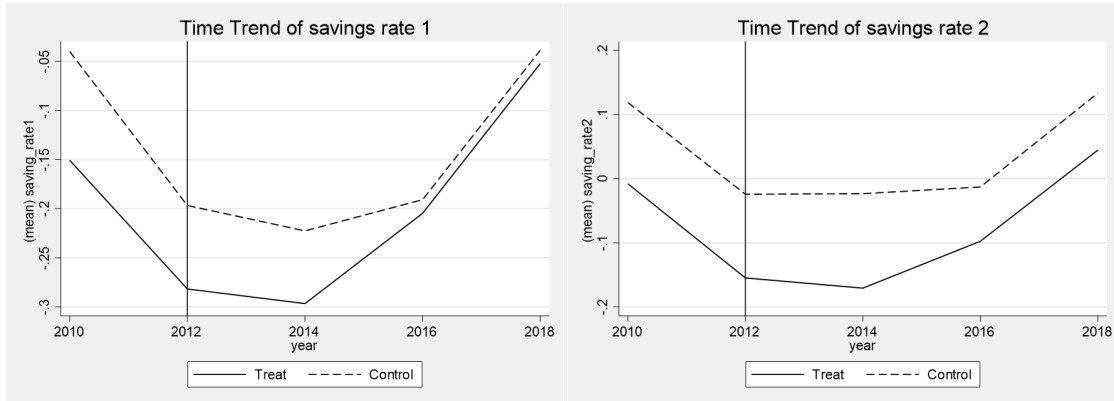

Fig. 3 Time Trend of Savings Rate

**Notes:** The left one uses the first definition of savings rate while the right one uses the second. A vertical line is set for the year 2012, for the reason that the policy announcement was made in 2013, and 2010 and 2012 are pre-treat periods.

---

[3] Admittedly, there are only two pre periods in the figure, which may raise questions about the validity of the parallel trend assumption. However, I believe that the existence of such a perfectly parallel line enables my study to progress convincingly.

## B. The Average Response to Delayed Retirement Announcement

According to Equation (1), the savings rate, calculated in two different ways, serves as the dependent variable. The results in Table 2 indicate that the announcement of the delayed retirement policy significantly increases the household savings rate for the treatment group. This finding holds valid even after controlling for household income, household savings, total assets, house value, household size, hospitalization rate, retirement rate, as well as individual and year fixed effects. In terms of the magnitude of the effect, the policy expectation leads to a 5 to 8% increase in the household savings rates for the treatment. This effect is not only statistically significant but also economically meaningful.

Table 2 The Average Response to Delayed Retirement Announcements: Baseline Regression

|  | Savings rate 1 | Savings rate 1 | Savings rate 2 | Savings rate 2 |
|---|---|---|---|---|
|  | (1) | (2) | (3) | (4) |
| Treat×Post | 0.0787** | 0.0732** | 0.0558** | 0.0558** |
|  | (0.0362) | (0.0329) | (0.0284) | (0.0284) |
| logarithm of household income |  | 0.746*** |  | 0.641*** |
|  |  | (0.0162) |  | (0.0139) |
| logarithm of household savings |  | 0.00550*** |  | 0.00272* |
|  |  | (0.00175) |  | (0.00154) |
| logarithm of total assets |  | -0.0499*** |  | -0.0496*** |
|  |  | (0.00802) |  | (0.00744) |
| household population size |  | -0.0874*** |  | -0.0767*** |
|  |  | (0.00930) |  | (0.00790) |
| hospitalization rate |  | -0.0973*** |  | -0.00681 |
|  |  | (0.0142) |  | (0.0121) |
| retirement rate |  | 0.142*** |  | 0.133*** |
|  |  | (0.0298) |  | (0.0258) |
| house value |  | -0.000322*** |  | -0.000286*** |
|  |  | (0.000119) |  | (0.000107) |
| Household FE | Yes | Yes | Yes | Yes |
| Year FE | Yes | Yes | Yes | Yes |
| Covariates | No | Yes | No | Yes |
| Observations | 13,939 | 13,254 | 13,215 | 13,215 |
| R-squared | 0.330 | 0.511 | 0.508 | 0.508 |

**Notes:** This table shows the average response to the delayed retirement announcement in the period from 2014 to 2018. Treat×Post is the variable indicating whether the treated households experiencing policy announcements. Robust standard errors are in parentheses and are clustered at the household level. *** P<0.01, ** P<0.05, and * P<0.1.

The estimated coefficients of covariates indicate that higher household income and a larger retirement ratio are associated with higher savings rates, which aligns with the theory of household consumption smoothing. Besides, larger household sizes and higher hospitalization rates are always associated with greater daily expenses, and the regression coefficients demonstrate their negative impact on household savings rates.

## C. The Dynamics of the Savings Rate Response

Since the announcement of the delayed retirement policy, various forms of news or rumors have been circulated through the media.[4] Additionally, people's expectations regarding this policy have also changed over time. Therefore, the impact of policy expectations on household savings rates may vary with the timing of the policy announcement. To identify the dynamic effects of the delayed retirement policy announcement on household savings rates, I conducted regressions based on Equation (2), and the results are presented in Table 3 and Figure 4.

Table 3 Dynamic Effects on Savings Rates

|  | Savings rate 1 | Savings rate 2 | Savings rate 1 | Savings rate 2 |
|---|---|---|---|---|
|  | (1) | (2) | (3) | (4) |
| Treat×2014 | 0.0113 | 0.0113 | 0.0990** | 0.0801* |
|  | (0.0581) | (0.0581) | (0.0482) | (0.0412) |
| Treat×2016 | 0.0771 | 0.0771 | 0.0961* | 0.0687 |
|  | (0.0591) | (0.0591) | (0.0510) | (0.0425) |
| Treat×2018 | 0.0645 | 0.0645 | 0.0419 | 0.0432 |
|  | (0.0572) | (0.0572) | (0.0493) | (0.0416) |
| Household FE | Yes | Yes | Yes | Yes |
| Year FE | Yes | Yes | Yes | Yes |
| Covariates | No | No | Yes | Yes |
| Observations | 13,939 | 13,939 | 13,254 | 13,215 |
| R-squared | 0.330 | 0.330 | 0.511 | 0.508 |

**Notes:** This table shows the dynamic effects on savings rates after the delayed retirement announcement (2013.11) in the period from 2014 to 2018. Treat×2014, Treat×2016 and Treat×2018 are the products of the treat dummy and year dummy. Robust standard errors are in parentheses and are clustered at the household level. *** P<0.01, ** P<0.05, and * P<0.1.

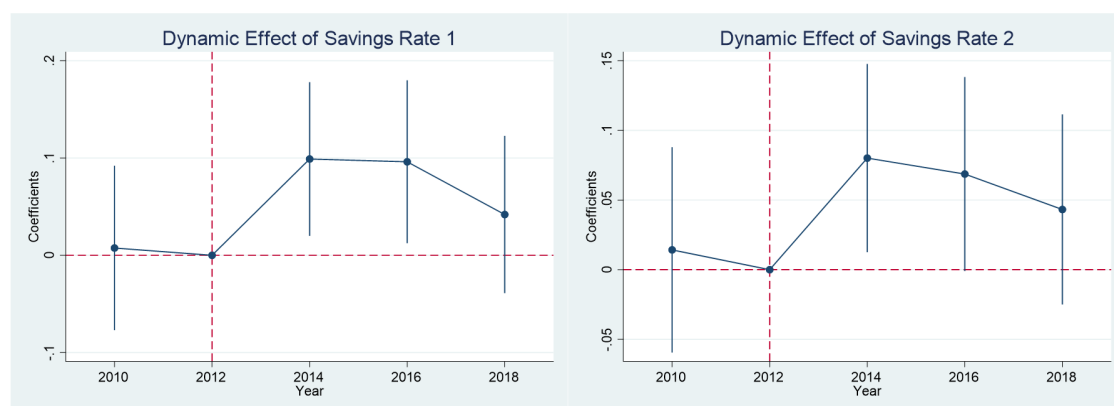

Fig. 4 Dynamic Effects on Savings Rate

**Notes:** This figure plots the dynamic effects on savings rates. The baseline samples are 2012 households. Results are from regression (3) and (4). The line perpendicular to the horizontal axis

---
[4] Which you can refer to "Baidu Search index".

indicates the 90% confidence interval.

As shown in Table 3 and Figure 4, the delayed retirement announcement can significantly increase household savings rates in period 0 (2014) and period 1 (2016), but has no significant impact in period 2 (2018). This may be attributed to the absence of specific policies during these years (2014-2018), leading to a weakening of people's expectations regarding the policy announcement. Additionally, it can be observed from the graph that the estimated coefficient for the second pre-policy period (2010) is not significantly different from zero. This further supports the validity of the parallel trend assumption.

# IV. Heterogeneity and Robustness

## A. Heterogeneity

The effect of delayed retirement announcement on the savings rates of the treated households may vary depending on household income and the employment status of family members. In this section, I will examine these two scenarios individually.

### 1. Household Income Differences

The households are categorized as "Above Average Income Households" or "Below Average Income Households" based on their income levels compared to the average income within their respective groups in 2014. Regression analyses are then performed separately for the treatment group and control group within each income category. The table below displays the results of these regressions.

Table 4 Households Savings Rates Response to Policy Announcement: Income Level Differences.

|  | Above Average Income Households | | Below Average Income Households | |
| --- | --- | --- | --- | --- |
|  | Savings rate 1 | Savings rate 2 | Savings rate 1 | Savings rate 2 |
|  | (1) | (2) | (3) | (4) |
| Treat×Post | 0.111** | 0.0574 | 0.0515 | 0.0583 |
|  | (0.0487) | (0.0422) | (0.0440) | (0.0381) |
| Household FE | Yes | Yes | Yes | Yes |
| Year FE | Yes | Yes | Yes | Yes |
| Covariates | Yes | Yes | Yes | Yes |
| Observations | 4,798 | 4,810 | 8,456 | 8,405 |
| R-squared | 0.469 | 0.470 | 0.523 | 0.520 |

**Notes:** This table shows the average effects on savings rates after the delayed retirement announcement in the period from 2014 to 2018, for two categories. Robust standard errors are in parentheses and are clustered at the household level. *** P<0.01, ** P<0.05, and * P<0.1.

According to the table, it can be observed that higher-income households tend to have higher savings rates compared to the other group. This could be attributed to their long-term financial planning and the potential concern about future financial constraints of pension insurance. So, they may anticipate a greater need to rely on their savings during retirement, motivating them to increase their savings rates. Meanwhile, the limited improvement in savings rates among lower-income households may be attributed to their higher proportion of daily living expenses relative to their income, resulting in a smaller elasticity of willingness to save.

**2. Employment Type Differences**

The type of employment may also affect the willingness to save. The study classifies households into "Self-employed" and "Employed by Others" based on the workplace of the household financial reporter. Again, I re-estimate the effect of the announcement. The table below displays the results of these regressions.

Table 5 Households Savings Rates Response to Policy Announcement: Employment Type Differences.

|  | Self-employed | | Employed by Others | |
| --- | --- | --- | --- | --- |
|  | Savings rate 1 | Savings rate 2 | Savings rate 1 | Savings rate 2 |
|  | (1) | (2) | (3) | (4) |
| Treat×Post | 0.0782* | 0.0746** | 0.0801** | 0.0574* |
|  | (0.0434) | (0.0376) | (0.0371) | (0.0326) |
| household FE | Yes | Yes | Yes | Yes |
| Year FE | Yes | Yes | Yes | Yes |
| Covariates | Yes | Yes | Yes | Yes |
| Observations | 5,563 | 5,529 | 7,374 | 7,344 |
| R-squared | 0.519 | 0.530 | 0.497 | 0.489 |

**Notes:** This table shows the average effects on savings rates after the delayed retirement announcement in the period from 2014 to 2018, for two categories. Robust standard errors are in parentheses and are clustered at the household level. *** P<0.01, ** P<0.05, and * P<0.1.

Policy expectations have a similar effect on household savings rates regardless of the employment type. Both experience an increase of approximately 7% to 8% in their savings rates.

## B. Robustness

**1. DID with a Continuous Treatment**

Relying solely on whether all household adult members have retired or not to determine the treatment and control groups will overlook the impact of the policy announcement's intensity. Let's consider an example of a three-member household: one with both spouses working and another with only one spouse working. The effect of the delayed retirement policy would differ between these two scenarios. To account for this, I introduce a proxy called "Policy Intensity," which measures the proportion of non-retired adults to all adults in the household. This variable captures the intensity of policy effects. Using the DID method, I re-estimate the regression coefficients.

Table 6 The Effect of Delayed Retirement Announcement on Different Policy Intensity Households

|  | Savings rate 1 | Savings rate 2 | Savings rate 1 | Savings rate 2 |
| --- | --- | --- | --- | --- |
|  | (1) | (2) | (3) | (4) |
| Policy Intensity×Post | 0.102*** | 0.0868*** | | |
| | (0.0359) | (0.0314) | | |
| Policy Intensity×2014 | | | 0.0913* | 0.0927** |
| | | | (0.0516) | (0.0450) |
| Policy Intensity×2016 | | | 0.156*** | 0.117*** |
| | | | (0.0526) | (0.0448) |
| Policy Intensity×2018 | | | 0.0975* | 0.0882** |
| | | | (0.0511) | (0.0438) |
| Household FE | Yes | Yes | Yes | Yes |
| Year FE | Yes | Yes | Yes | Yes |
| Covariates | Yes | Yes | Yes | Yes |
| Observations | 13,254 | 13,215 | 13,254 | 13,215 |
| R-squared | 0.511 | 0.508 | 0.511 | 0.508 |

**Notes:** This table shows both the average and dynamic effects on savings rates after the delayed retirement announcement in the period from 2014 to 2018. "Policy Intensity" is $\%FW_i$ in section II. Robust standard errors are in parentheses and are clustered at the household level. *** P<0.01, ** P<0.05, and * P<0.1.

The table reveals that the delayed retirement announcement has a stronger impact on households with higher policy intensity, with coefficients of 0.102 and 0.087, significant at the 1% level. The average proportion of non-retired household members among all adults in the treatment group is 0.84. This implies an average 8% increase in savings rates for the treated group due to the expected effect of the delayed retirement policy (based on a savings rate of 1). Additionally, the estimation of dynamic effects aligns with the findings in Section III Part C, showing an initial increase followed by a decline in policy effects.

## 2. Placebo Test

To demonstrate the real influence of the delayed retirement announcement on the savings rates of non-retired households, this study uses a random assignment approach to re-assign treatment and control groups. The key explanatory variable's coefficient estimates and standard errors are calculated with 1000 random assignments. The t-values of the coefficients are reported in the following figure.

The distribution of t-values for the coefficient of interest closely approximates a normal distribution with a mean of 0. Only a small proportion of estimates show a significant deviation from 0. This finding provides additional evidence that the policy announcement indeed has a significant effect on the households affected by it.

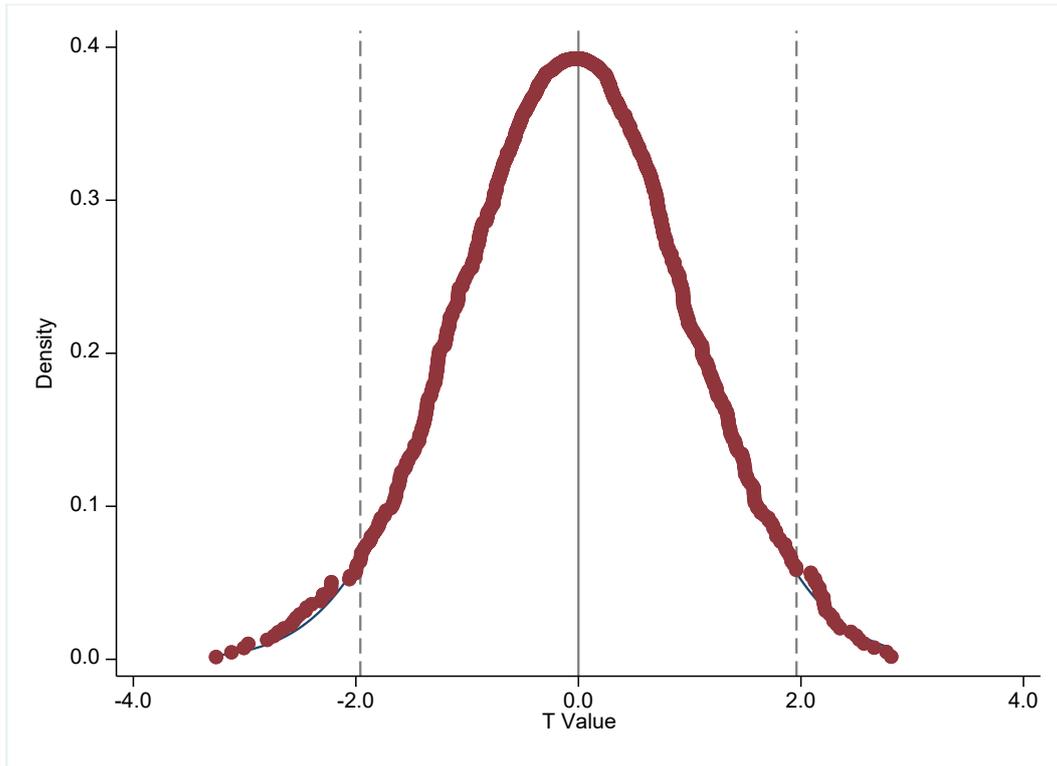

Fig. 5 2.　Placebo Test: Distribution of the T Values of the Coefficient of Interest

**Notes:** The estimates of T values in this figure come from 1000 regressions, each with its random grouping.

## V. Mechanisms

In this section, I examine several household expenditures as the dependent variable and explain the reasons behind the decrease in expenditure and increase in savings rates among the treatment group households. As indicated in the table, the reduction in expenditure among the households affected by the policy announcement is mainly observed in medical expenses, other expenses, and mortgage expenditures. While these channels may contribute to the increase in savings rates, they are not directly attributable to the delayed retirement policy announcement. A more confounding finding is that the changes in daily expenses vary a lot, with some categories showing increases and others showing decreases. I believe that the changes in savings rates and various expenditure categories form a comprehensive system that is indeed related to the policy announcement of delayed retirement. However, there remains a distinct gap between the policy itself and the specific decision-making processes concerning household economic matters.

Table 7 The Announcement Effect: A Channel Analysis of The Impact on Household Expenditures

| | Food Expenditure | Dress Expenditure | Housing Expenditure | Daily Expenditure | Medical Expenditure |
|---|---|---|---|---|---|
| | (1) | (2) | (3) | (4) | (5) |
| Treat×Post | -558.7 | 563.9*** | 1,591* | 5,080** | -2,538*** |
| | (597.8) | (130.2) | (900.8) | (2,312) | (752.3) |
| Household FE | Yes | Yes | Yes | Yes | Yes |
| Year FE | Yes | Yes | Yes | Yes | Yes |
| Covariates | Yes | Yes | Yes | Yes | Yes |
| Observations | 15,155 | 15,133 | 15,098 | 15,016 | 15,228 |
| R-squared | 0.578 | 0.509 | 0.290 | 0.242 | 0.335 |

(Continued)

| | Transportation & Communication Expenditure | Education & Entertainment Expenditure | Other Expenditure | Mortgage Expenditure | Commercial Insurance Expenditure |
|---|---|---|---|---|---|
| | (6) | (7) | (8) | (9) | (10) |
| Treat×Post | 806.3*** | 501.2 | -783.4*** | -1,177 | 1,223*** |
| | (191.4) | (417.6) | (284.9) | (827.3) | (197.6) |
| Household FE | Yes | Yes | Yes | Yes | Yes |
| Year FE | Yes | Yes | Yes | Yes | Yes |
| Covariates | Yes | Yes | Yes | Yes | Yes |
| Observations | 15,072 | 15,194 | 15,214 | 15,298 | 15,223 |
| R-squared | 0.601 | 0.467 | 0.234 | 0.260 | 0.471 |

**Notes:** This table shows the average expenditure changes after the delayed retirement announcement in the period from 2014 to 2018. Treat×Post is the variable indicating whether the treated households experiencing policy announcements. Robust standard errors are in parentheses and are clustered at the household level. *** P<0.01, ** P<0.05, and * P<0.1.

The introduction of the delayed retirement policy is related to the financial situation of the government, especially the financial pressure on pensions. With the increase in the aging population and the gradual decline in the proportion of the young population due to the long-term one-child policy, an earlier retirement age will bring an excessive number of retirees and correspondingly create a labor shortage. Therefore, the introduction of the delayed retirement policy for one reason aims to reduce the government's pension burden. Each household also faces the trade-off between saving and consumption and should accumulate wealth during their working years to prepare for retirement. The implementation of the delayed retirement policy necessitates families to engage in long-term financial planning to some extent. They consider the uncertainty of post-retirement living expenses and pensions, leading to an increased emphasis on saving. Additionally, if the delayed retirement policy is indeed put into effect, rational households will anticipate potential reductions in future pension amounts or a delay in receiving pensions. This motivation prompts them to save more to bridge the economic gap during retirement.

One piece of evidence is the increase in household expenditure on commercial insurance in the treatment group, as shown in the figure below. Since the announcement of the delayed retirement in 2013, households' spending on commercial insurance has increased each year, and this effect is progressively cumulative. This reflects the fact that households are planning more for future consumption and corroborates with the findings of increased household savings rates confirmed by this article.

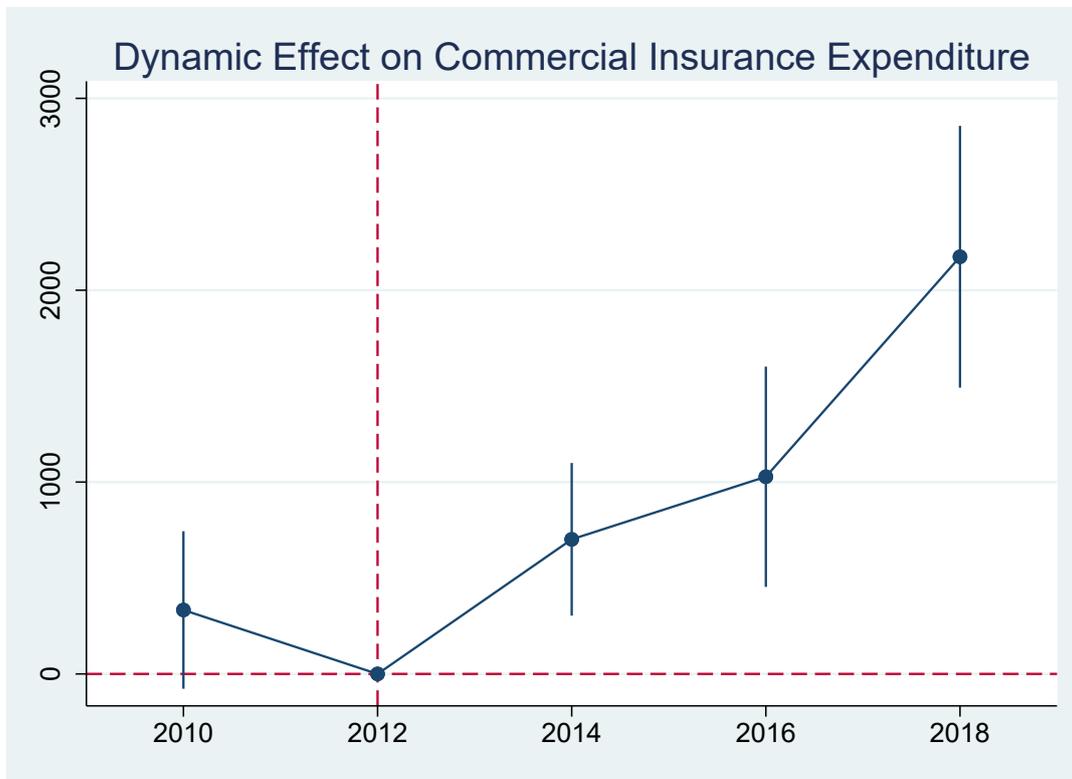

Fig. 6 Dynamic Effect on Commercial Insurance Expenditure

**Notes:** This figure plots the dynamic effect on commercial insurance expenditure. The baseline samples are 2012 households. The line perpendicular to the horizontal axis indicates the 95% confidence interval.

## VI. Conclusion

This article utilizes panel data from China Family Panel Studies (CFPS) and focuses on savings rates as the dependent variable. The Difference-in-Differences (DID) method is employed to estimate the impact of the announcement of the delayed retirement policy in 2013 on households that have members not yet retired. Furthermore, the dynamic effects of the policy and alternative estimation methods are explored to obtain robust conclusions.

The findings reveal that the announcement of the delayed retirement policy significantly increases the savings rates of the affected households, with an average effect of 7% to 8% increase. The dynamic effects of this policy announcement vary over time, with the most pronounced effects observed in the initial two periods after the policy announcement, followed by a gradual reduction. This may be attributed to the lack of specific policies, which could affect people's expectations of the policy's feasibility. The heterogeneity analysis shows that higher-income households are more likely to be impacted by policy expectations, leading to a more significant increase in their savings rate. Robustness further confirms the finding.

Regarding the mechanism analysis, it is challenging to determine the exact reasons for the increased savings rates in the treatment group by solely examining household expenditures, as expenditure changes are influenced by multiple factors, making it difficult to separate the effects of the delayed retirement policy announcement from other social factors. However, a closer examination reveals that families may anticipate a reduction in future pension amounts or a delay in pension receipt. This helps explain the increase in their savings rate. Additionally, the study finds that households in the treatment group have increased their expenditure on commercial insurance, providing additional evidence for the conclusion.

The main contribution of this study lies in the use of micro-level data to observe the decision outcomes of households under policy expectations, while previous research has mainly focused on policy analysis at the national or governmental level. Moreover, as the delayed retirement policy has not been implemented but only announced, it holds special academic value. This study demonstrates that policy expectations can have a similar influence on household decisions as the actual policy implementation, as indicated by the title "Policy Expectation Counts." If the delayed retirement policy is implemented in the future, the findings of this study can serve as a reference for assessing its policy effects, allowing for a more accurate estimation of the policy impact by considering the effects of policy expectations.

# REFERENCES

**[1] Agarwal, Sumit, and Wenlan Qian.** 2014. "Consumption and Debt Response to Unanticipated Income Shocks: Evidence from a Natural Experiment in Singapore." American Economic Review, 104 (12): 4205-30.

**[2] Callaway, B., Goodman-Bacon, A., & Sant'Anna, P. H.** 2021. Difference-in-differences with continuous treatment. arXiv preprint arXiv:2107.02637.

**[3] Can Lui, Chen Ling, and Hong Zhou.** 2019 "The Announcement of Increasing-Retirement-Age Policy and Changes in Savings Rates of Urban Households." Finance & Trade Economics, Vol. 40, No.4,2019: 130-145.

**[4] Chen, Yi, Ziying Fan, Xiaomin Gu, and Li-An Zhou.** 2020. "Arrival of Young Talent: The Send-Down Movement and Rural Education in China." American Economic Review, 110 (11): 3393-3430.

**[5] Moser, Petra, and Alessandra Voena.** 2012. "Compulsory Licensing: Evidence from the Trading with the Enemy Act." American Economic Review, 102 (1): 396-427.

**[6] Liu, Z.** 2022. Delayed Retirement, Individual Account Adjustment and Sustainable Operation of Urban Employee Medical Insurance Funds in China. Individual Account Adjustment and Sustainable Operation of Urban Employee Medical Insurance Funds in China (November 14, 2022).